\documentclass[sigconf]{acmart}
\usepackage{balance}
\usepackage{latexsym}
\usepackage{multicol}
\usepackage{multirow}
\usepackage{booktabs}
\usepackage{algorithm}
\usepackage{algorithmic}
\usepackage{fancyhdr}
\usepackage{subfigure}
\usepackage{makecell}
\usepackage{enumitem}
\AtBeginDocument{%
  \providecommand\BibTeX{{%
    \normalfont B\kern-0.5em{\scshape i\kern-0.25em b}\kern-0.8em\TeX}}}

\copyrightyear{2025}
\acmYear{2025}
\setcopyright{acmlicensed}\acmConference[WWW '25]{Proceedings of the ACM
Web Conference 2025}{April 28-May 2, 2025}{Sydney, NSW, Australia}
\acmBooktitle{Proceedings of the ACM Web Conference 2025 (WWW '25), April
28-May 2, 2025, Sydney, NSW, Australia}
\acmDOI{10.1145/3696410.3714829}
\acmISBN{979-8-4007-1274-6/25/04}

\begin{document}


\pagestyle{fancy}
\fancyhf{}
\fancyhead[L]{\conferenceaddress}
\fancyhead[R]{\shortauthors}

\title{Unleashing the Potential of Two-Tower Models: Diffusion-Based Cross-Interaction for Large-Scale Matching}

\author{Yihan Wang}
\email{wangyihan05@kuaishou.com}
\affiliation{%
  \institution{Kuaishou Technology}
  \city{Beijing}
  \country{China}
}

\author{Fei Xiong}
\email{xiong_info@163.com}
\affiliation{%
  \institution{Unaffiliated}
  \city{Beijing}
  \country{China}
}

\author{Zhexin Han}
\email{hanzhexin03@kuaishou.com}
\affiliation{%
  \institution{Kuaishou Technology}
  \city{Beijing}
  \country{China}
}

\author{Qi Song}
\email{songqi@kuaishou.com}
\affiliation{%
  \institution{Kuaishou Technology}
  \city{Beijing}
  \country{China}
}

\author{Kaiqiao Zhan}
\email{zhankaiqiao@kuaishou.com}
\affiliation{%
  \institution{Kuaishou Technology}
  \city{Beijing}
  \country{China}
}

\author{Ben Wang}
\authornote{Corresponding Author}
\email{wangben@kuaishou.com}
\affiliation{%
  \institution{Kuaishou Technology}
  \city{Beijing}
  \country{China}
}
\renewcommand{\shortauthors}{Yihan Wang et al.}
\newcommand{\conferenceaddress}{WWW '25, April 28-May 2, 2025, Sydney, NSW, Australia}


\begin{abstract}
Two-tower models are widely adopted in the industrial-scale matching stage across a broad range of application domains, such as content recommendations, advertisement systems, and search engines. This model efficiently handles large-scale candidate item screening by separating user and item representations. However, the decoupling network also leads to a neglect of potential information interaction between the user and item representations. Current state-of-the-art (SOTA) approaches include adding a shallow fully connected layer(i.e., COLD), which is limited by performance and can only be used in the ranking stage. For performance considerations, another approach attempts to capture historical positive interaction information from the other tower by regarding them as the input features(i.e., DAT). Later research showed that the gains achieved by this method are still limited because of lacking the guidance on the next user intent. To address the aforementioned challenges, we propose a "cross-interaction decoupling architecture" within our matching paradigm. This user-tower architecture leverages a diffusion module to reconstruct the next positive intention representation and employs a mixed-attention module to facilitate comprehensive cross-interaction. During the next positive intention generation, we further enhance the accuracy of its reconstruction by explicitly extracting the temporal drift within user behavior sequences. Experiments on two real-world datasets and one industrial dataset demonstrate that our method outperforms the SOTA two-tower models significantly, and our diffusion approach outperforms other generative models in reconstructing item representations.
\end{abstract}

\ccsdesc[500]{Information systems~Recommender systems}
\ccsdesc[500]{Information systems~Information retrieval}

\keywords{Candidate Matching, Diffusion Models, Embedding-based Retrieval}



\maketitle
\thispagestyle{empty}

\section{Introduction}
\label{sec:Introduction}

\begin{figure}[!t]
  \centering
  \subfigure[]{
    \includegraphics[width=0.3\linewidth]{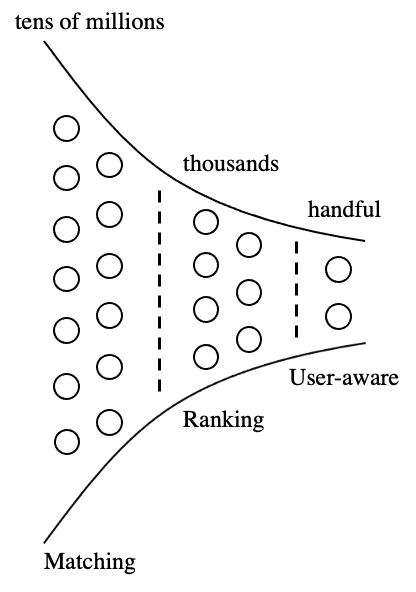}
    \label{fig:two_tower(a)}
  }
  \subfigure[]{
    \includegraphics[width=0.63\linewidth]{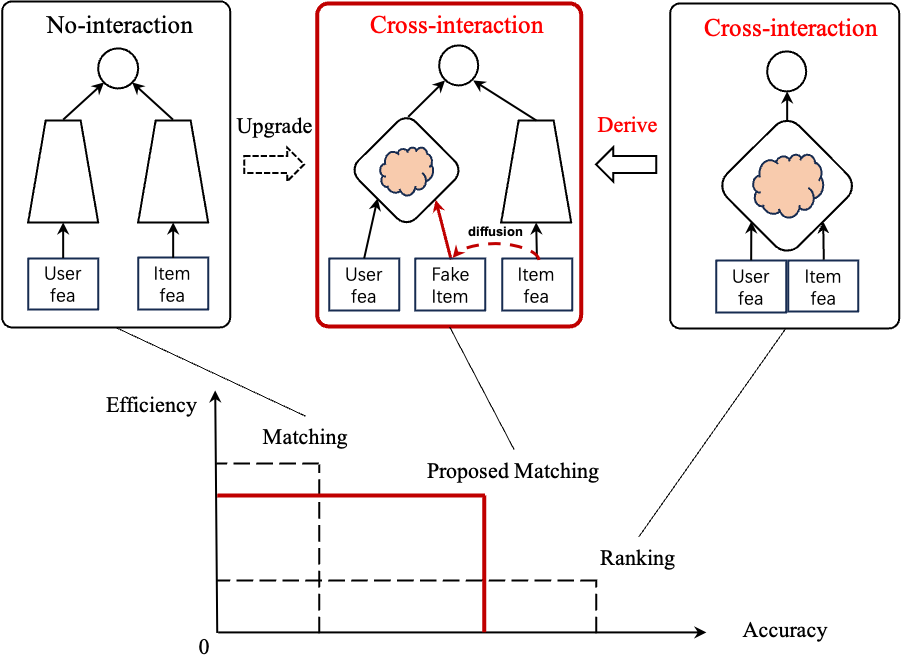}
    \label{fig:two_tower(b)}
  }
  \caption{Real-world two-stage recommender system. (a)The two-stage architecture involves matching, which scores a large number of items, and ranking, which further refines the scoring for a smaller subset. (b) Intuitive view for accuracy and efficiency of matching and ranking method, where the proposed matching method is derived from ranking and optimized to a cross-interaction architecture.}
\end{figure}

Recommender systems aim to enhance user experience and business value by suggesting items of interest and driving user engagement and satisfaction. In the industry scenario, a two-stage recommender system, as shown in Figure \ref{fig:two_tower(a)}, is extensively used for providing users with personalized content with strict latency. The first stage is called the matching stage, which narrows down the potential set of candidates from a large corpus. The second stage, known as the ranking stage ~\cite{casmos, explore_rec}, selects the final results that the user might be interested in.

The matching stage is a critical phase of recommender systems where filters out the irrelevant candidates from billions of corpus quickly. Due to the high accuracy and low latency requirements of the matching models, two-tower models ~\cite{dssm, mix-neg, DAT, simplex} become a primary paradigm for candidate matching and support for efficient top-k retrieval ~\cite{beyond}. The Two-tower model consists of two separate towers, one tower processes all the information about the query (user, context), while the other tower processes information about the candidates. The outputs of two towers are low-dimensional embeddings, which are then multiplied for scoring candidate items.

Since the two-tower models trained independently, they cannot leverage cross-features or interactions between user and item features until the very end, which is referred to as "Late Interaction" ~\cite{colbert}. Recent research on fetching the interactive signals can primarily be categorized into two approaches. One method transforms the two-tower architecture into the single-tower structure by adding a shallow fully connected layer (i.e., COLD ~\cite{cold} and FSCD ~\cite{towardsab}), but the efficiency is still constrained and can only be used in the ranking phase. The other method attempts to augment the embedding input of each tower with a vector that captures historical positive interaction information from the other tower(i.e., DAT ~\cite{DAT}), recent research shows that the gains are still limited ~\cite{inttower} because of lacking the guidance on the next user positive intent. Current SOTA approaches are difficult to balance model effectiveness and inference efficiency. Figure \ref{fig:two_tower(b)} describes the aforementioned models from the perspective of inference efficiency and prediction accuracy. 

To tackle the trade-off between efficiency and accuracy, we propose a generative cross-interaction decoupling architecture of the matching paradigm, named Unleashing the Potential of \textbf{T}wo-\textbf{T}ower Models: \textbf{Diff}usion-Based Cross-Interaction for Large-Scale Matching (\textbf{T2Diff}). T2Diff has exceeded the limits of the two-tower architecture by extracting user-item cross features with the guidance of target item restored by the diffusion module. Considering the performance issues caused by the large-scale corpus in the matching phase, instead of the single-tower structure, we employ a generative method that reconstructs the user's positive interactions contained in the item tower through a diffusion model in the user tower. To model the interactions between user and item features sufficiently, a mixed-attention module is introduced to enhance the user's positive interaction from the other tower. This mixed-attention module extracts user representation more accurately by interacting with the item information and the user's historical behavior sequence. The main contributions of this paper are as follows:
\begin{itemize}[leftmargin=*]
\item We propose a new matching paradigm named T2Diff which is a generative cross-interaction decoupling architecture that emphasizes information interactions and unleashes the potential of two-tower model with high accuracy and low latency.
\item T2Diff introduces two key innovations: 1) a generative module to reconstruct user's next positive intention by applying diffusion-based model, and 2) a mixed-attention mechanism ~\cite{Transfomer, DIN} to address the challenge of the "Late Interaction" by facilitating more complex and enriched user-item feature interactions at the foundational level of the model architecture.
\item T2Diff not only outperforms the baselines on both two real-world datasets and one industrial dataset, but also demonstrates great inferences efficiency.
\end{itemize}

\section{Related Works}
\label{sec:Related Works}

\noindent\textbf{Embedding-based Retrieval} (EBR): A technique that uses embeddings to represent users and items, converting the retrieval problem into a nearest neighbor (NN) search problem in the embedding space ~\cite{Faiss, scann}. EBR models are widely applied in the matching stage ~\cite{EBR-fb}, which selects a list of candidates from a large corpus based on the user’s historical behavior. Typically, EBR models consist of two parallel deep neural networks for learning the encoding of the users and items, which are trained separately and also known as two-tower model ~\cite{dssm, mix-neg, youtube}. This architecture has the advantages of high throughput and low latency, while the ability to capture the interactive signals between user and item representations is limited. To mitigate the problem, DAT ~\cite{DAT} introduces a adaptive-mimic mechanism which customizes an augmented vector for each user and item, compensating for the lack of interactive signals. However, later research ~\cite{inttower} shows that the gain of only introducing an augmented vector as the input features is limited. Therefore, T2Diff leverages the mixed-attention module to extract high-order feature interactions and user historical behaviors with the target representations generated by diffusion module.

\noindent\textbf{Session-based Recommendation and Interests Drift.} Feng $et\ al.$ \cite{dsin} have observed that user behaviors within each session exhibit a high degree of homogeneity, yet they tend to drift across different sessions. Zhou $et\ al.$ \cite{DIEN} have discovered that the accuracy of predicting the Click-Through Rate (CTR) is significantly enhanced when the predictions are aligned with the trend of interests drift.

\noindent\textbf{The Application of Generative Model in Sequential Recommendation.} Although
traditional sequential models, such as SASRec \cite{Sasrec}, Mamba4Rec \cite{Mamba4rec} have demonstrated satisfactory performance, the emergence of generative models has revealed a new and promising direction. VAEs ~\cite{beta-VAE, ContrastVAE, STOSA} have been utilized to learn a latent space representation of items and users, from which new sequences can be generated. However,  these kind of generative models might oversimplify the data distribution, leading to a loss of information and potentially less accurate representations. Diffusion models have made remarkable success in many fields, including recommender systems~\cite{diffRec, DiffuRec, cam_ae, giffcf}, natural language processing ~\cite{beta-VAE, proDiff, diffusionModel}, and computer vision ~\cite{DDPM, DDIM, LDM}. DiffuRec ~\cite{DiffuRec} made the first attempt to apply diffusion modeling to SR and adopted a single embedding to fetch a user’s multiple interests due to its ability of distribution generation and diversity representation. While VAEs and diffusion models applied in computer vision ~\cite{beta-VAE, proDiff, diffusionModel} typically rely on a Kullback-Leibler divergence loss[KL-loss] to measure the difference between the learned latent distribution and a prior distribution (often a Gaussian), DiffuRec opts for a cross-entropy loss during the process of reconstructing the target item. In order to restore item representation stably and accurately, T2Diff adopts a diffusion module with Kullback-Leibler divergence loss[KL-loss]. This module can accurately reconstruct the target item with low latency, providing a solid foundation for capturing cross-information within the two-tower structure.

\section{Preliminary}
\label{sec:Preliminary}
In this section, we briefly introduce the diffusion models as preliminary knowledge.

\subsection{Diffusion Models}
\label{sec:Diffusion Models}
Diffusion models can be divided into two stages, diffusion process and reverse process. Fundamentally, Diffusion Models work by destroying training data through the successive addition of Gaussian noise in diffusion process, and then learning to recover the data by reversing this noising process in reverse process.

In the diffusion process, the diffusion models add the Gaussian noise successively to the original representations $x_0$ via a Markov Chain (i.e., $x_0 \rightarrow x_1 \rightarrow ... \rightarrow x_T $) as follows:
\begin{equation}
    \begin{aligned}
        q(x_t|x_{t-1}) = \mathcal{N}(x_t;\sqrt{1-\beta_t} x_{t-1},\beta_t I)
    \end{aligned}
    \label{eq1}
\end{equation}
where $\mathcal{N}(x;\mu,\sigma^2)$ is a Gaussian distribution with mean $\mu$ and variance $\sigma^2$. $\beta_t$ represents the amplitude of added Gaussian noise, with higher values of $\beta_t$ indicating a higher level of introduced noise. $I$ is the identity matrix.

We can go in a closed form from the input data $x_0$ to $x_T$ in a tractable way and the posterior probability can be defined as:
\begin{equation}
    \begin{aligned}
        q(x_{1:T}|x_0) = \prod_{t=1}^T q(x_t|x_{t-1})
    \end{aligned}
    \label{eq2}
\end{equation}

According to DDPM ~\cite{DDPM}, with the help of reparameterization trick, we can find that the posterior $q(x_r|x_{0})$ obey a Gaussian distribution. Let $\alpha_r = 1-\beta_r$ and $\bar{\alpha}_r = \Pi_{i=1}^r \alpha_i$, then the Equation \ref{eq2} can be rewritten as 
\begin{equation}
    \begin{aligned}
        q(x_r|x_{0}) = \mathcal{N}(x_r;\sqrt{\bar{\alpha}_r} x_{0},(1-\alpha_r) I)
    \end{aligned}
    \label{eq_reparam}
\end{equation}

In the reverse process, we gradually denoise from the standard Gaussian representation $x_T$ and approximate the real representation $x_0$ $(i.e. x_T\rightarrow x_{T-1} \rightarrow ... \rightarrow x_0)$ in an iterative way. Specially, given the current restored representation $x_t$ and the original representation $x_0$, the next representation $x_{t-1}$ can be calculated as follows:
\begin{equation}
    \begin{aligned}
        p(x_{t-1}|x_t,x_0) = \mathcal{N}(x_{t-1};\tilde{\mu}_t(x_t,x_0),\tilde{\beta}_t I) 
    \end{aligned}
    \label{eq3}
\end{equation}
\begin{equation}
    \begin{aligned}
        \tilde{\mu}_t(x_t,x_0) = \frac{\sqrt{\bar{\alpha}_{t-1}}\beta_t}{1-\bar{\alpha}_t}x_0 + \frac{\sqrt{\alpha_t}(1-\bar{\alpha}_{t-1})}{1-\bar{\alpha}_t}x_t 
    \end{aligned}
    \label{eq4}
\end{equation}
\begin{equation}
    \begin{aligned}
        \tilde{\beta}_t = \frac{1-\bar{\alpha}_{t-1}}{1-\bar{\alpha}_t}\beta_t
    \end{aligned}
    \label{eq5}
\end{equation}

However, the original representation $x_0$ is always unknown in the reverse process, thus requiring a deep neural network to estimate $x_0$. The reverse process is optimized by minimizing the following variational lower bound (VLB).
\begin{equation}
    \begin{aligned}
        L_{VLB} &= E_{q(x_1|x_0)}[logp_{\theta}(x_0|x_1)] - D_{KL}(q(x_T|x_0)||p_{\theta}(x_T))   \\
        &  \quad  - \sum_{t=2}^T E_{q(x_t|x_0)}[D_{KL}(q(x_{t-1}|x_t,x_0)||p_{\theta}(x_{t-1}|x_t))]  \\
        & = L_0- L_T- \sum_{t=2}^T L_{t-1}
    \end{aligned}
    \label{eq:VLB}
\end{equation}
where $p_{\theta}(x_t) = \mathcal{N}(x_t;0,I)$ and $D_{KL}(\cdot)$ is the KL divergence.

Each KL divergence term in \( L_{\text{VLB}} \), with the exception of \( L_0 \), involves the comparison of two Gaussian distributions. As such, these terms can be analytically computed in closed form. The term \( L_T \) is constant during the training process, making it inconsequential for optimization. This is because the distribution \( q \) lacks trainable parameters and \( x_T \) is simply Gaussian noise. For modeling \( L_0 \), Ho $et\ al.$ utilize a separate discrete decoder derived from \( N \). Following ~\cite{DDPM}, $L_{VLB}$ can be simplified as a Gaussian noise learning process, which can denoted as
\begin{equation}
    \begin{aligned}
        L_{simple} = E_{t\in [1,T],x_0,\epsilon_t}\left[||\epsilon_t-\epsilon_{\theta}(x_t,t)||^2 \right]
    \end{aligned}
    \label{eq7}
\end{equation}
where $\epsilon \sim \mathcal{N}(0,I)$ is sampled from a standard Gaussian distribution, and $\epsilon_{\theta}(\cdot)$ represents an \textbf{Estimator} that can be learned by a deep neural network.

\section{Method}
\label{sec:Method}
\begin{figure*}[h]
  \centering
  \subfigure[]{
    \includegraphics[width=0.28\linewidth]{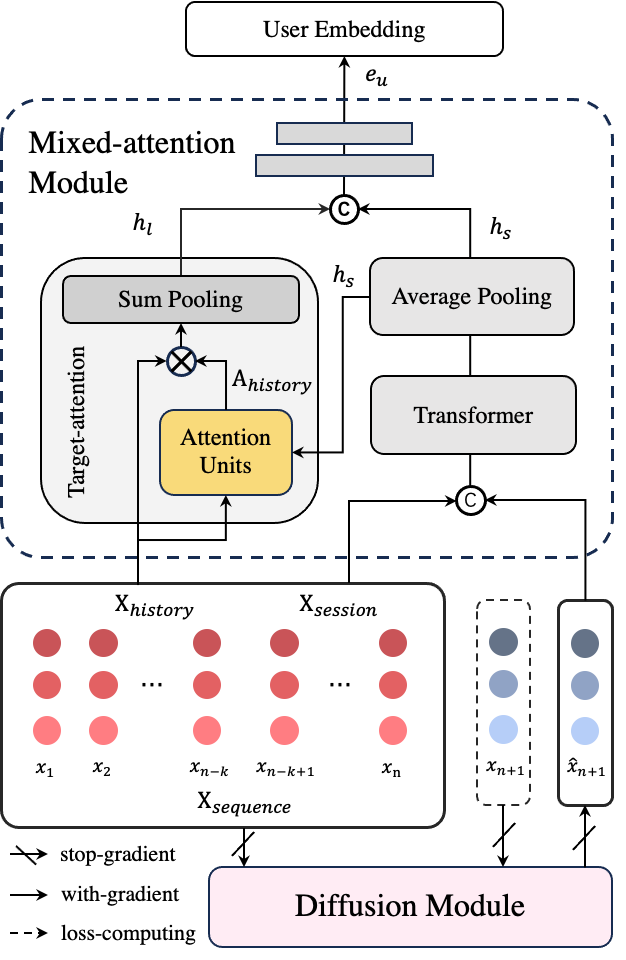}
    \label{fig:overview(a)}
  }
  \subfigure[]{
    \includegraphics[width=0.65\linewidth]{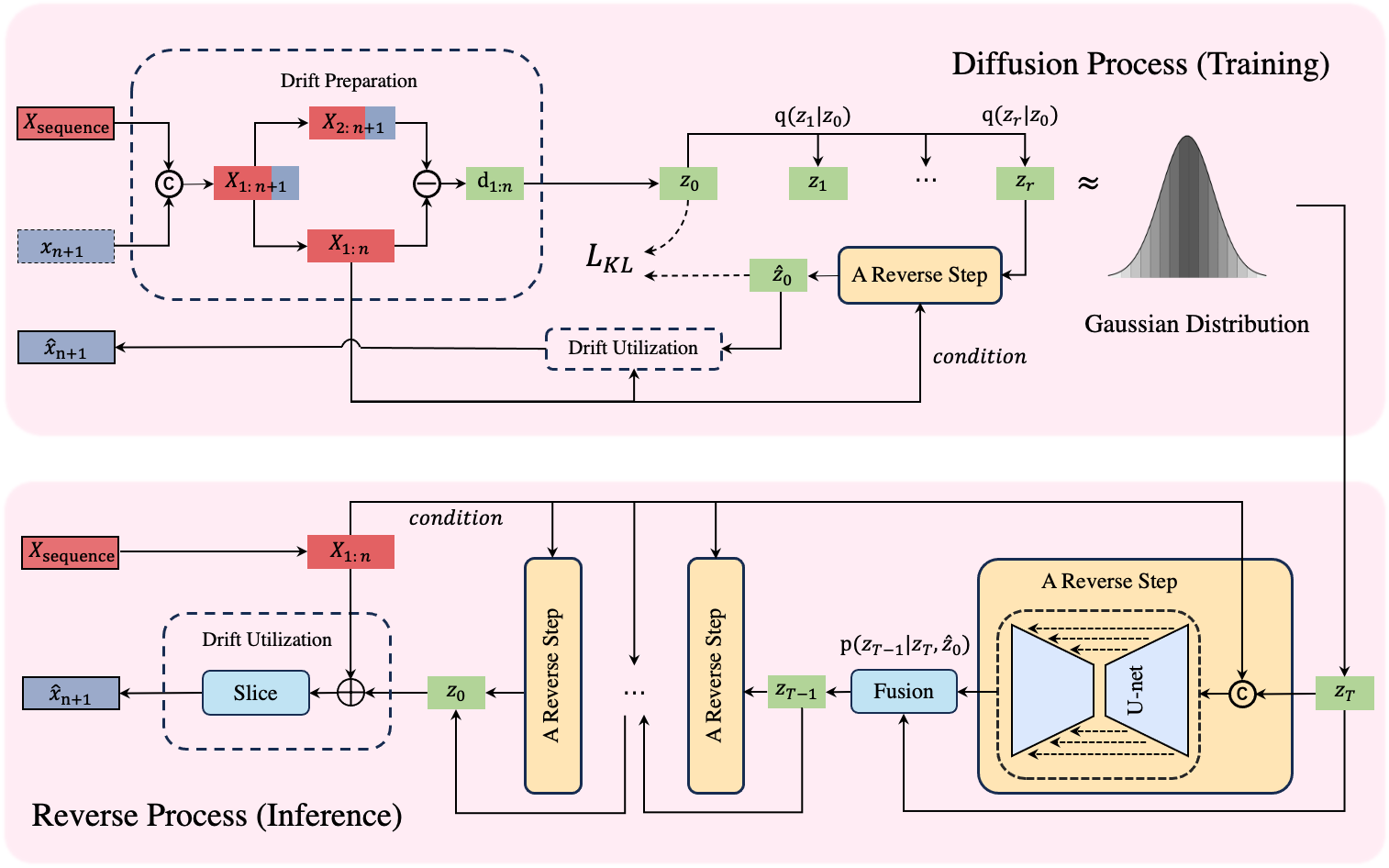}
    \label{fig:overview(b)}
  }
  \caption{(a): The main architecture of T2Diff includes a mixed-attention module and a diffusion module. The forward slash character used to connect the diffusion module and embedding layer indicates stop-gradient. (b): The details of the diffusion module, which adopts different processes for training and inference. For each reverse step in the diffusion module, we utilize a Unet approximator.}
\end{figure*}

In this section, we first introduce the notation and background related to T2Diff. We then detail the framework of our model, which consists of a diffusion module and a mixed-attention module, as shown in the Figure \ref{fig:overview(a)}.

\subsection{Notations and Problem Formulation}
\label{sec:Notations}
Suppose that we have a set of users $\mathcal{U}$ and a set of items $\mathcal{M}$. We collect the behavior sequence of each user and denote it as $X_{sequence} \in M$. We tell each behavior of user $u \in \mathcal{U}$ as $x^u_j$, where j represents the $j\text{-}th$ item of the behavior sequence. For each user, suppose that we have $n$ historical behaviors, then index $j \in \{1, 2, \cdots, n+1\}$ and $X_{sequence} = [x_1, x_2, \cdots, x_{n}]$. Building upon the concept presented in \cite{dsin}, we aim to achieve a more refined modeling of user behavior sequences by dividing them into two distinct parts based on the time intervals that separate each action. Specifically, we divide the ordered behavior sequence into the current session with recent $k$ interacted behaviors denoted by $X_{session} = [x_{n-k+1}, \cdots, x_{n}]$ and historical behaviors denoted as $X_{history} = [x_1, x_2, \cdots, x_{n-k}]$. We believe that the user's behaviors within the most recent session are temporally continuous, and reflect the user's nearest intentions. Finally, the most important point, we unleash the two-tower model's potential by introducing the predicted next positive behavior $\hat{x}_{n+1}$ from the true one $x_{n+1}$. 

Embedding-based retrieval (EBR) methods encode user and item features into embeddings by two independent deep neural networks. The relevance of item $\mathcal{M}$ to user $\mathcal{U}$ is determined based on the distance (most commonly, inner product) between user embedding $e_u$ and item embedding $e_i$. 

Our proposed T2Diff has two main parts: 1) A diffusion module designed to identify drift in interests between adjacent behaviors during the training phase, and to reintroduce next behavior during the inference stage. 2) A session-based mixed-attention module that extracts current interests from the latest session and the predicted next behavior by applying a self-attention module and fetching historical interests with a target-attention mechanism. The combination of these two components enables a full cross-interaction between the user's behavior sequence and the next behavior.

\subsection{Diffusion Module}
\label{sec:Diffusion Module}
Referring to the bottom of Figure \ref{fig:overview(a)}, the input of the diffusion module is the complete user's behaviors $X_{sequence}$ and the next positive behavior $x_{n+1}$, which fed into the diffusion process and transformed into the standard Gaussian distribution. The output is the predicted next positive behavior $\hat{x}_{n+1}$ restored from a sample of standard Gaussian distribution, which is expected to be the same as $x_{n+1}$ in the reverse process. 
\begin{equation}
    \begin{aligned}
        \hat{x}_{n+1} = \text{Diffusion}(X_{session}, x_{n+1})
    \end{aligned}
    \label{eq8}
\end{equation}
Regarding the drift preparation step illustrated on the upper left of Figure \ref{fig:overview(b)}, assuming there are $n$ user historical behaviors and $1$ next behavior, concatenating these behaviors in time series we can get $X_{1: n+1}$. One necessary step is to obtain the drift between adjacent behaviors. We employ sliding windows to derive $X_{1: n}$ and $X_{2: n+1}$ separately, from those we calculate the element-wise subtraction of $X_{2: n+1}$ from $X_{1: n}$, resulting in $d_{1: n}$ or $z_0$, to which we add noise. 
\begin{equation}
    \begin{aligned}
        X_{1:n+1} = \text{concat}([X_{sequence}, x_{n+1}])
    \end{aligned}
    \label{eq6}
\end{equation}
\begin{equation}
    \begin{aligned}
        z_0 = d_{1:n} = X_{2: n+1} - X_{1: n}
    \end{aligned}
    \label{eq9}
\end{equation}
We believe that diffusion and reverse from drift between adjacent behaviors is much easier than from the original user behavior sequence $X_{1: n}$. The effectiveness of this approach will be demonstrated by experiments in Section \ref{sec:Ablation experiments}. 

\subsubsection{$\textbf{Diffusion Process}$}
\label{sec:Diffusion Process}
During the training process, we randomly add $r$ steps of Gaussian noise to a batch of data, which can be achieved by 1 step from $q(\cdot)$, according to Equation \ref{eq_reparam}. The step-index $r$ is randomly selected from Uniform distribution $[1, T]$, where $T$ is the upper limit of the diffusion step. We have devised an exponential noise schedule $\beta$ to introduce noise incrementally, showcasing its novelty in Section \ref{sec:Diffusion Model Hyperparam}. The procedure of a single diffusion step can be represented by Equation \ref{eq10}-\ref{eqz_r}:
\begin{equation}
    \begin{aligned}
        r \sim \text{Uniform}(\text{\{1,...,T\}})
    \end{aligned}
    \label{eq10}
\end{equation}
\begin{equation}
    \begin{aligned}
        \beta_r = a\cdot e^{br}
    \end{aligned}
    \label{eq_beta}
\end{equation}
\begin{equation}
    \begin{aligned}
        z_r &= \sqrt{\bar{\alpha}_r}z_0+\sqrt{1-\bar{\alpha}_r}\epsilon \\
            &\sim q(z_r|z_0)
    \end{aligned}
    \label{eqz_r}
\end{equation}
where a, b, T are hyper-parameters, $\epsilon \sim \mathcal{N}(0, I)$. Next, we choose U-Net as the backbone of our approximator to recover its unbiased estimation $\hat{z}_0$. The traditional U-Net architecture comprises an encoder, a decoder, and skip connections, facilitating the generation of an output that keeps dimensional congruence with the input. One of the salient advantages of utilizing U-Net lies in its convolutional kernels, which can capture user interest drift over time. This feature significantly augments the model's capacity for discerning intricate user interest patterns within the original input sequence and effectively reconstructing them from a noised input, denoted as $z_r$.

Notably, conditional diffusion models incorporate additional information as input, such as the class label $c$. In our context, the user's original behavior sequence $X_{1: n}$ is rich in information regarding interest drift and is accessible during both the training and inference phases. Consequently, $X_{1: n}$ is employed as a conditional factor to direct the reverse direction of the approximator.
\begin{equation}
    \begin{aligned}
        \hat{z}_0 = \text{U-Net}(\text{concat}([z_r, X_{1: n}]))
    \end{aligned}
    \label{eq11}
\end{equation}

\subsubsection{$\textbf{Reverse Process}$}
\label{sec:Reverse Process}
Following the application of Diffusion Models in Computer Vision ~\cite{DDPM,DDIM,LDM}, even if we start reversing next behavior from a sample of standard Gaussian noise $z_T \sim \mathcal{N}(0, I)$, it is possible to recover the drift between the last behavior $x_n$ and the next behavior $x_{n+1}$ with the guidance of original user behavior $X_{1: n}$. In each reverse step, as shown in Equation \ref{eq11} and Equation \ref{eq12}, we obtain $\hat{z}_0$ from the U-Net approximator with learnable parameters, denoted as $f_{\theta}$. Then we combine $\hat{z}_0$ with $z_t$ to get $z_{t-1}$ via $p(\cdot)$. After applying the reparameterization trick, a single reverse step presents as follows: 
\begin{align}
    \hat{z}_0   &= f_{\theta}(z_0 | z_t, X_{1: n})  \label{eq12} \\
    z_{t-1} &= \text{Fusion}(z_t, \hat{z}_0) 	\notag \\
            &= \tilde{\mu}_t(z_t,\hat{z}_0)+\tilde{\beta}_t\epsilon' \\ 
            &\sim p(z_{t-1}|z_t, \hat{z}_0)  \notag
\end{align}

where $\epsilon' \sim \mathcal{N}(0, I)$. After T reverse steps, the output $z_0$ is still an intermediate one. Therefore, in the drift utilization step, we add it to $X_{1: n}$, and obtain the last behavior as predicted next behavior $\hat{x}_{n+1}$, as shown on the lower left of Figure \ref{fig:overview(b)}.
\begin{equation}
    \begin{aligned}
        \hat{x}_{n+1} = \text{Slice}(z_0+X_{1: n})
    \end{aligned}
    \label{eq14}
\end{equation}
where $\text{Slice}(x) = x[-1]$. The reverse process of the diffusion module is illustrated in Algorithm \ref{alg:reverse_phase}.

\begin{algorithm}[!h]
    \caption{Diffusion Process(Training)}
    \label{alg:diffusion_phase}
    \begin{algorithmic}[1]
    \STATE \textbf{Inputs:}
    \STATE $\quad \text{User Historical Sequence: } X_{sequence} \text{ or } X_{1: n}$
    \STATE $\quad \text{Next Behavior: } x_{n+1}$
    \STATE $X_{1: n+1} = \text{concat}([X_{sequence}, x_{n+1}])$
    \STATE $z_0 = X_{2: n+1} - X_{1: n} \text{ // drift preparation}$
    \STATE $r \sim \text{Uniform}(\text{\{1,...,T\}})$
    \STATE $z_r \sim q(z_r|z_0)$
    \STATE $\hat{z}_0 = \text{U-Net}(\text{concat}([z_r, X_{1: n}]))$
    \STATE $\text{parameter update}: L_{KL}(\hat{z}_0, z_0)$
    \STATE $\hat{x}_{n+1} = \text{Slice}(\hat{z}_0+X_{1: n}) \text{ // drift utilization}$
    \RETURN{$\hat{x}_{n+1}$}
    \end{algorithmic}
\end{algorithm}

\begin{algorithm}[h]
    \caption{Reverse Process (Inference)}
    \label{alg:reverse_phase}
    \begin{algorithmic}[1]
    \STATE \textbf{Inputs:}
    \STATE $\quad \text{User Historical Sequence: } X_{sequence} \text{ or } X_{1: n}$
    \STATE $\quad \text{Gaussian Sampling: } z_T \sim \mathcal{N}(0,I)$
    \FOR{$t=T,...,1$}
        \STATE $\hat{z}_0 = \text{U-Net}(\text{concat}([z_t, X_{1: n}]))$
        \STATE $z_{t-1} = \text{Fusion}(z_t, \hat{z}_0)$
    \ENDFOR
    \STATE $\hat{x}_{n+1} = \text{Slice}(z_0 + X_{1: n}) \text{ // drift utilization}$
    \RETURN{$\hat{x}_{n+1}$}
    \end{algorithmic}
\end{algorithm}

\subsection{Mixed-attention Module}
\label{sec:Mixed-attention Module}
\begin{figure}[h]
  \centering
  \includegraphics[width=0.6\linewidth]{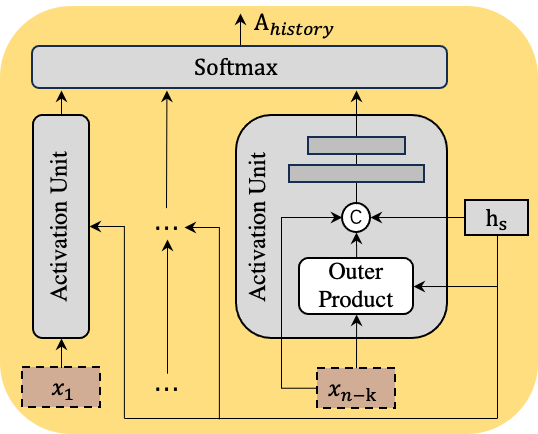}
  \caption{The detailed implementation of activation units used for target-attention.}
  \label{fig:activation}
\end{figure}

To overcome the issue of "Late Interaction" in the two-tower model, we propose a mixed-attention mechanism that facilitates intricate feature interactions by engaging multi-layer user representations with the reconstructed user's recent positive item representation obtained by the diffusion module in Section \ref{sec:Diffusion Module}. In the realm of short-video recommendation, user consumption behaviors demonstrate temporal continuity. We consider that the last session contains the user's recent positive intention, and to to enhance the cross-interactions between historical sequences and the next positive item representation, we concatenate $X_{session}$ and $\hat{x}_{n+1}$ along the temporal dimension. In our approach, we deploy the encoder component of the transformer architecture ~\cite{Transfomer} and average pooling to generate current interests embedding $h_s$ for "Early Interaction". 
\begin{equation}
    \begin{aligned}
        h_s = \text{avg}(\text{Transfomer}(\text{concat}([X_{session}, \hat{x}_{n+1}])))
    \end{aligned}
    \label{eq15}
\end{equation}
To further exploit the benefit of cross-interaction, following ~\cite{DIN}, we use $h_s$ as guidance to extract similar information from the user's historical behaviors $X_{history}$. In the activation units, the historical behavior embeddings $X_{history}$, the current interests embedding $h_s$, and their outer product are provided as the inputs for generating attention weights $A_{history}$, as illustrated in Figure \ref{fig:activation}. Finally, $h_t$ and $h_s$ will collectively determine the user embedding $e_u$.
\begin{equation}
    \begin{aligned}
        a_j = \frac{\text{FFN}(\text{concat}([x_j, x_j-h_s, x_j*h_s, h_s]))}{\sum_{i=1}^{n-k}\text{FFN}(\text{concat}([x_i, x_i-h_s, x_i*h_s, h_s]))}
    \end{aligned}
    \label{eq16}
\end{equation}
\begin{equation}
    \begin{aligned}
        h_l = f(h_s, [x_1, x_2,\cdots,x_{n-k}])
        = \sum_{j=1}^{n-k} a_jx_j
    \end{aligned}
    \label{eq17}
\end{equation}
\begin{equation}
    \begin{aligned}
        e_u = \text{FFN}(\text{concat}([h_l, h_s)])
    \end{aligned}
    \label{eq18}
\end{equation}
where $a_j$ is the $j\text{-}th$ element of $A_{history}$. Considering the temporal dependencies within a session and the correlations of behavioral patterns across sessions, we introduce the time lag between the target behavior and historical behaviors as a critical feature.

\subsection{Model Optimization}
\label{sec:Model Optimization}

In each diffusion step, we derive $\hat{z}_0$ directly from $z_r$, where $\hat{z}_0$ and $z_0$ both represent the mean of the distribution by reparameterization. Therefore, the simplified version of $L_{VLB}$ from Equation \ref{eq:VLB} can be rewritten as follows, denoted as $L_{KL}$,
\begin{equation}
    \begin{aligned}
        L_{KL} = E_{r\in [1,T],x_0,\mu_r}\left[||\mu_r-\mu_{\theta}(z_r,r)||^2 \right]
    \end{aligned}
    \label{eq19}
\end{equation}
where $\mu_r$ and $z_r$ represent the noise added during the diffusion process at step r and the results after adding noise, respectively, $\mu_{\theta}$ denotes the estimator with parameters $\theta$.

With the help of $L_{KL}$, we can reduce the difference between $z_0$ and $\hat{z}_0$ and renew the parameters in the approximator by gradient descent. The diffusion process of the diffusion module is illustrated in Algorithm \ref{alg:diffusion_phase}. 

Following the general principles of loss function in the recommender systems, a softmax loss $L_{TOWER}$ is utilized to bring the user embedding $e_u$ close to the target item embedding $e_i$ while far away from the remaining irrelevant item embeddings $e_{m\in \mathcal{M}}$, which is denoted as
\begin{equation}
    \begin{aligned}
        L_{TOWER} = -\log \frac{exp(e_u\cdot e_i)}{\Sigma_{m\in \mathcal{M}}exp(e_u \cdot e_m)}
    \end{aligned}
\end{equation}
Enabled by the loss function $L_{TOWER}$, the sparse embedding table undergoes thorough training, thereby establishing a robust foundation for the diffusion process training. The total loss can be denoted as 
\begin{equation}
    \begin{aligned}
        L_{TOTAL} = L_{TOWER} + \lambda L_{KL}.
    \end{aligned}
\end{equation}
where $\lambda$ is a hyper-parameter, usually set to 1 or 10. Because the optimization direction of the approximator within the diffusion module is inconsistent with that of traditional recommender systems, which will easily lead to a situation where gradients counteract each other, so we employ a stop-gradient mechanism to isolate the gradient updates of the diffusion module, effectively enhancing the optimization efficiency of both the approximator and the tower parameters, as shown on the bottom of Figure \ref{fig:overview(a)}.

\section{Experiments}
\label{sec:Experiments}
In this section, we conduct experiments on two benchmark datasets frequently utilized in the research community and a large-scale industrial dataset consisting of one million users from an online short-video platform. In detail, we try to answer the following questions:

$\bullet$ How effective is the proposed method compared to other SOTA two-tower sequential recommendation models? \textbf{Q1} 

$\bullet$ What are the distinct contributions of the individual modules within our proposed model, including the diffusion module for the reconstruction of the next behavior and the mixed-attention module that facilitate efficient cross-interaction? \textbf{Q2} 

$\bullet$ How should the hyper-parameters of the diffusion module be selected to strike a balance between accuracy and efficiency, particularly with respect to the noise generation schedule $\beta$ and the upper limit of diffusion step $T$? \textbf{Q3} 

\subsection{Offline Evaluation}
\label{sec:Offline Evaluation}
\subsubsection{Datasets}
\label{sec:Datasets}
We conduct extensive offline experiments over two public datasets, KuaiRand and ML-1M. The number of users, items and interactions are displayed in the Table \ref{table_dataset}.

\begin{table}[!h]
  \caption{Statistics of the datasets}
  \begin{tabular}{cccc}
    \toprule
    Dataset& \#Users & \#Items & \#Interactions\\
    \midrule
    KuaiRand & 25,828 & 108,025  & 6,492,153\\
    ML-1M  & 6,040 & 3,648 & 643,979\\
  \bottomrule
\end{tabular}
\label{table_dataset}
\end{table}

\textbf{KuaiRand} ~\cite{kuairand} is a publicly available dataset collected from the logs of the recommender system in Kuaishou. Following ~\cite{divide_conquer}, we extract user-items interactions from main recommendation scenario where the "tab" field equals one and treat clicked items as relevant to user. 

\textbf{ML-1M} ~\cite{ml_1m} consists of over one million anonymous ratings of about 3648 movies made by 6040 MovieLens users, which is widely used in other sequential-aware methods, such as Caser~\cite{Caser} and SASRec~\cite{Sasrec}. All the movies watched by a user are considered relevant. 

For both datasets, we focus on positive samples for training purposes, specifically clicksed items in KuaiRand and watched items in ML-1M. Additionally, we exclude users with interaction frequencies below five to ensure a more robust training dataset.

\begin{table*}[!t]
\centering
\caption{Performance comparison on ML-1M and KuaiRand with other SOTA methods. The best results of all methods are highlighted in bold font and the best results of the baselines are \underline{underlined}. 'Improvement' is the relative improvement against the best baseline performance. All the performance gains are statistically significant at $p < 0.05$. 'Params' denotes parameters. 'Infer Time' is the inference time consumption per sample, test on 5 Tesla T4 GPUs.}
\begin{tabular}{ccccc|cccccc}
\hline
Dataset                 & \multicolumn{4}{c}{ML\_1M}                           & \multicolumn{4}{c}{KuaiRand($\times\text{1e-1}$)}   &  \multirow{3}{*}{\makecell[c]{Params \\(MB)}}  & \multirow{3}{*}{\makecell[c]{Infer Time \\(ms)} }                   \\
\cline{1-5} \cline{6-9}
\multirow{2}{*}{Method} & \multicolumn{2}{c}{Recall} & \multicolumn{2}{c}{MRR} & \multicolumn{2}{c}{Recall} & \multicolumn{2}{c}{MRR} & \\
\cline{2-5} \cline{6-9}
                        & 2            & 20          & 2          & 20         &     10         & 100            &       10      &    100 &       \\
                        \hline
SASRec                  & 0.05779      & 0.15564     & 0.04515    & 0.05717     & 0.03796              & 0.21192              & 0.01631              & 0.01991    &   0.126  & 0.56 \\
Caser                   & 0.05808      & 0.15709     & 0.04259    & 0.05769    & 0.04076              & 0.19544              & 0.01653              & 0.01981       &  0.182 & 0.64 \\
GRU4Rec                 & 0.06070      & 0.16564     & 0.04503    & 0.05753    & 0.03909              & 0.21304              & 0.01492              & 0.01880      &  0.119 & 0.60 \\
Bert4Rec                & 0.06272      & 0.20559     & 0.04589    & 0.06532    & 0.04041              & 0.19902              & 0.01465              & 0.01898       &  0.126 & 0.57 \\
DAT+                     & 0.05988      & \underline{0.25785}     & 0.04506    & \underline{0.07140}    &   0.04334           &      0.20760       &     0.01320        &     0.01624    &  0.127  & 0.54  \\
Mamba4Rec            & \underline{0.06919}       & 0.20581     & \underline{0.05262}    & 0.06888    &   \underline{0.05219}       &      \underline{0.22800}      &     \underline{0.02031}        &     \underline{0.02378}  & 0.131 & 1.11  \\
\hline
ContrastVAE             & 0.06023      & 0.15203     & 0.04436    & 0.05764     & 0.04195              & 0.21155              & 0.01605              & 0.02024    &   0.126 & 0.63 \\
STOSA                   & 0.05905      & 0.16008     & 0.04454    & 0.05705     & 0.03745              & 0.21365              & 0.01678              & 0.01991      &   0.244   & 0.56 \\
DiffuRec                & 0.06076      & 0.14971     & 0.04288    & 0.05469    & 0.04349              & 0.22614              & 0.01708              & 0.02028     &   0.146  & 0.65\\
\hline
Mixed-attention           & 0.06308      & 0.26017     & 0.04826    & 0.07452    &        0.04502      &      0.23990       &      0.01735       &     0.02310   &  0.126 & 0.59\\
W/O DP                  & 0.07233      & 0.25198     & 0.05683    & 0.07996    &        0.05014      &      0.24200       &      0.01694       &      0.02142     &  0.187 & 0.68 \\
\textbf{T2Diff(Ours)}                    & \textbf{0.07738}    & \textbf{0.27727}     & \textbf{0.06076}   &\textbf{0.08730}    &  \textbf{0.05505}            &   \textbf{0.25294}          &     \textbf{0.02106}        &     \textbf{0.02475}    & 0.187 & 0.68 \\
\hline
Improvement             & +11.84\%      & +7.53\%     & +15.47\%    & +22.27\%    &        +5.48\%      &      +10.94\%       &      $+3.69\%$       &    $+4.08\%$   &  —— &  ——\\
\hline
\end{tabular}
\label{table_compare}
\end{table*}

\subsubsection{Evaluation Metrics and Baselines}
\label{sec:Evaluation Metrics and Baselines}

To emulate real-world scenarios, for each method, a candidate set of $\mathcal{K}$ most relevant items is generated. 
Then, we utilize $Recall@K$ and Mean Reciprocal $Rank(MRR)@K$ to evaluate the offline effectiveness, which are widely applied in recommender systems.
K is set to 2 and 20 for ML-1M dataset and 10 and 100 for KuaiRand dataset. 

We compare it with the following SOTA baselines:

$\bullet$ \textbf{SASRec} ~\cite{Sasrec} introduces self-attention mechanism to quickly generate user embedding.

$\bullet$ \textbf{Caser} ~\cite{Caser} uses convolutional layers and max-pooling operation to capture local and global patterns in the sequence.

$\bullet$ \textbf{GRU4Rec} ~\cite{GRU4Rec} employs multiple GRU units to effectively capture the complex relationships in a user historical behavior sequence.

$\bullet$ \textbf{Bert4Rec} ~\cite{Bert4Rec} introduces the Cloze objective and bidirectional transformer structure to accomplish target prediction.

$\bullet$ \textbf{ContrastVAE} ~\cite{ContrastVAE} employs a two-branched VAE framework guided by ContrastELBO to address the challenge for sparsity of user-item interations.

$\bullet$ \textbf{STOSA} ~\cite{STOSA} devises a novel Wasserstein Self-Attention module to characterize item-item position-wise similarity in the sequence.

$\bullet$ \textbf{DiffuRec} ~\cite{DiffuRec} employs diffusion models to accomplish item representation construction and uncertainty injection for sequence recommendation.

$\bullet$ \textbf{DAT+} ~\cite{DAT} integrates a Adaptive-Mimic Mechanism (AMM) to mitigate the lack of information interaction. In the experiment setting, we employ AMM base on regular two-tower architecture.

$\bullet$ \textbf{Mamba4Rec} ~\cite{Mamba4rec} firstly leverages the power of selective SSMs for efficient sequential recommendation to address the dilemma of recommendation performance and inference efficiency.

In summary, SASRec, Caser, GRU4Rec, Bert4Rec and Mamba4Rec are representative examples of traditional two-tower models. DAT is distinguished by its item-augmented approach within the same architectural framework. Furthermore, ContrastVAE, STOSA, and DiffuRec are recognized as generative models that innovate upon the two-tower paradigm. 

\subsubsection{Comparisons with SOTA. (Q1)}
\label{sec:Comparisons with SOTA}
Due to the sheer volume of data in the KuaiRand dataset, it is challenging for the model to rank the target item within a smaller candidate set, which consequently results in relatively poorer overall performance when compared to the ML-1M dataset. To simplify the presentation, the results on the KuaiRand dataset will be multiplied by 10. The detailed quantitative comparison results is shown in Table \ref{table_compare}.

Among all traditional two-tower recommendation baselines, DAT+ outperforms others on some metrics, suggesting that it is beneficial to introduce additional interactions for two-tower model. However, DAT ignores the temporal relationship within the user historical behaviors, which limits the extent of performance improvement. In contrast, our proposed T2Diff effectively reconstructs target item representation by accounting for the temporal drift within user sequences, thereby achieving superior performance.

ContrastVAE and STOSA both utilize the Variational AutoEncoder (VAE) framework to model user behavior sequences. However, these models rely on two distinct embedding representations to capture the mean and variance, complicating the optimization process. DiffuRec utilizes diffusion model to introduce target information, but often struggles to achieve good performance due to ignoring the temporal relationship between target items and the users' historical sequence of actions.
 
In comparison to the best baseline, our proposed T2Diff demonstrates a significant enhancement in recall and MRR, specifically by $11.84\%$ and $22.27\%$, respectively, on the ML-1M dataset, and by $10.94\%$ and $4.08\%$, respectively, on the KuaiRand dataset.
It demonstrates that reconstructing target item is an effective strategy for addressing the "Late Interaction" problem, which significantly enhances the model's performance and elevates the target item to a desirable rank within the candidate set.

Additionally, we compare the computational complexity of our proposed T2Diff with other SOTA methods. As demonstrated in Table \ref{table_compare}, T2Diff achieves superior performance without a substantial increase in parameters. Furthermore, diffusion models are often associated with longer inference times, which can limit their applicability in real-world recommendation scenarios. To address this, we also calculate the inference time of various models, as presented in Table \ref{table_compare}. Compared to other SOTA methods, T2Diff maintains stable performance while keeping competitive time complexity.

\begin{figure}[b]
  \centering  \includegraphics[width=0.85\linewidth]{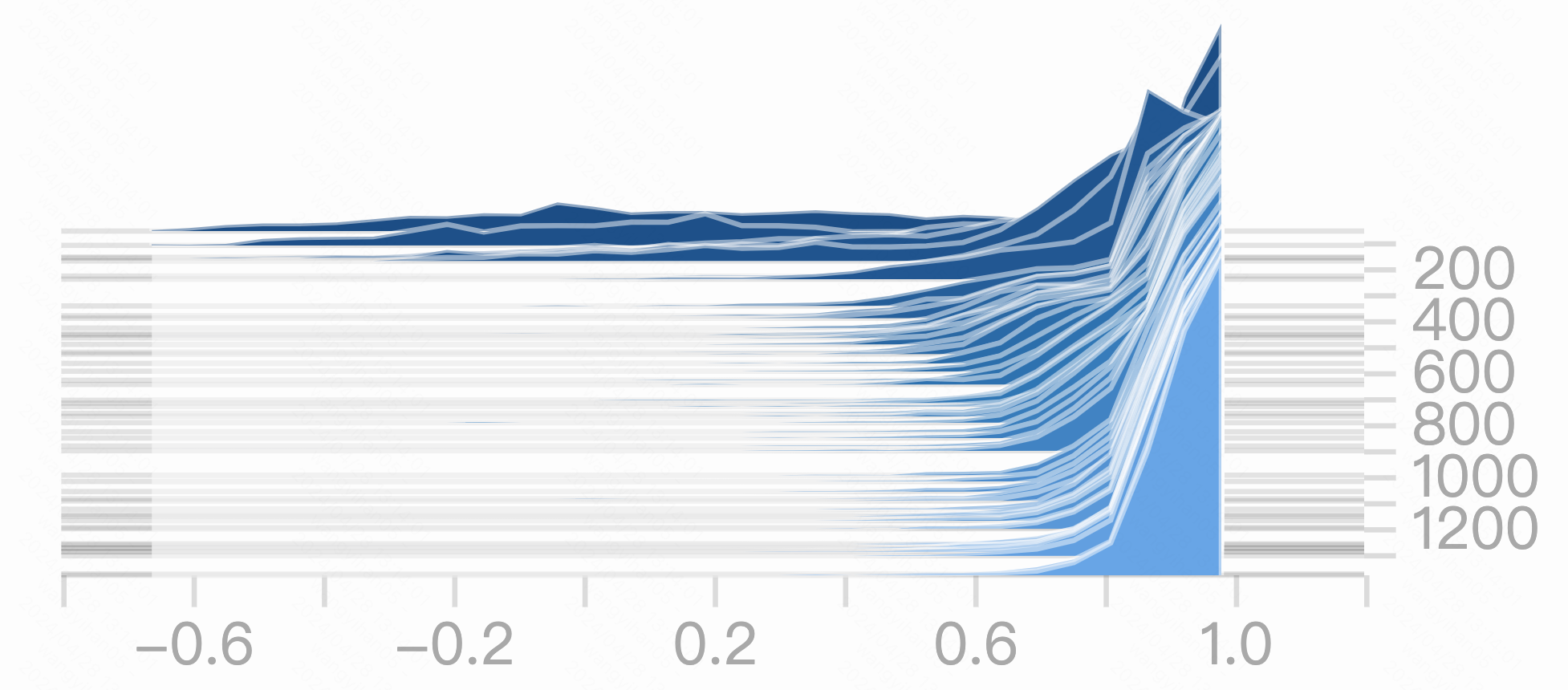}
  \caption{This figure delineates the distribution of similarities between $z_0$ and $\hat{z}_0$ across the diffusion process. The horizontal axis represents cosine similarity, while the vertical axis corresponds to the number of iterations.}
  \label{fig_similarity}
\end{figure}

\subsubsection{Ablation experiments. (Q2)}
\label{sec:Ablation experiments}
In this section, we will delineate the specific contributions of each module. Table \ref{table_compare} shows the ablation results on both datasets. Our model surpasses all baselines with the mere inclusion of the mixed-attention module in several metrics. Furthermore, by incorporating diffusion module, our model achieve significant improvements on both datasets, with recall rates increasing by $22.67\%$ on ML-1M, and by $25.90\%$ on KuaiRand, respectively.

Furthermore, the importance of fetching user interest drift can be evaluated by removing the drift preparation (DP) step. As shown in Table \ref{table_compare}, our experimental results demonstrate that the inclusion of the DP step significantly improved our model's performance on both datasets. These findings provide further evidence to support the notion that modeling user interest drift is vital for accurate prediction of user's preferred items. 

To further substantiate the corrective impact of the diffusion module during the diffusion and subsequent reverse processes, we track the cosine similarity between $z_0$ and $\hat{z}_0$ throughout the reverse process. As depicted in Figure \ref{fig_similarity}, the diffusion module demonstrated satisfactory performance in reversing $z_r$ back to $z_0$ after adequate amount of iterations.

\subsubsection{Verification of Hyper-parameters Selection for the Diffusion Module. (Q3)}
\label{sec:Diffusion Model Hyperparam}
To enhance the effectiveness of the Diffusion module, we undertake an investigation of the diffusion process by replacing various noise injection methods and optimizing the number of steps.
The detailed description of our experiments is provided below.

\textbf{Validation of $\beta$ design.} We conduct a comprehensive analysis of the impact of various noise generation methods. Specifically, we carry out a comparative study among three different approaches, called linear schedule, logarithmic schedule, and exponential schedule. 
As shown in Table \ref{table_noise_level}, our finding reveals that the exponential schedule, which is applied in our model, outperforms the popular linear schedule utilized in previous DiffuRec ~\cite{DiffuRec} studies. This approach can better satisfy the needs of the diffusion model for uniformly perturbing the inputs at each step, leading to improved performance and reliability of our results. 
\begin{table}[h]
\centering
\caption{Ablation studies of different methods to improve the noise level in diffusion process on ML-1M dataset.}
\resizebox{0.48 \textwidth}{!}{
\begin{tabular}{ccccc}
\hline
    $\beta$ schedule    & recall@2     & recall@20    & mrr@2 & mrr@20 \\
\hline
linear          & 0.07686         & 0.27519           & 0.05950  & 0.08677  \\
log                & 0.07256           & 0.25151          & 0.05767  & 0.08137   \\
\textbf{exp(ours)} & \textbf{0.07738} & \textbf{0.27727} & \textbf{0.06076}  & \textbf{0.08730}  \\
\hline
\end{tabular}
}
\label{table_noise_level}
\end{table}

\textbf{Validation of the maximum step in diffusion module.} We conduct an ablation study to ascertain the optimal diffusion step $T$ for our model. Experiments were conducted with $T$ set to 10, 50, and 200. As shown in Table \ref{table_step}, an increased diffusion step yielded progressively better performance for our proposed T2Diff model. Notably, while a diffusion step of 200 yielded the optimal MRR, further increments in $T$ dose not proportionally enhance performance over the $T=50$. Furthermore, elevating the diffusion step to 200 from 50 significantly amplified the inference time per sample by 238\%, which could impede the practical utility of T2Diff in industrial settings. Consequently, we have elected to set the diffusion step count at 50 for both industrial applications and all subsequent experiments detailed in this paper.

\begin{table}[h]
\centering
\caption{Ablation studies of maximum steps in diffusion module on ML-1M dataset.}
\resizebox{0.48 \textwidth}{!}{
\begin{tabular}{cccccc}
\hline
 steps & recall@2     & recall@20     & mrr@2         & mrr@20 & \makecell[c]{Infer Time \\(ms)}   \\
 \hline
10      & 0.0769     & 0.2724     & 0.0592    & 0.0852   & 0.22      \\
50(ours)   & \textbf{0.0774}    & \textbf{0.2773}     & 0.0608     & 0.0873   & 0.68    \\
200 & 0.0771 & 0.2715 & \textbf{0.0623} & \textbf{0.0885} & 2.30\\
\hline
\end{tabular}
}
\label{table_step}
\end{table}

\subsection{Live A/B Experiment} 
\label{sec:Live A/B Experiment}
\begin{figure}[h]
  \centering
  \includegraphics[width=0.75\linewidth]{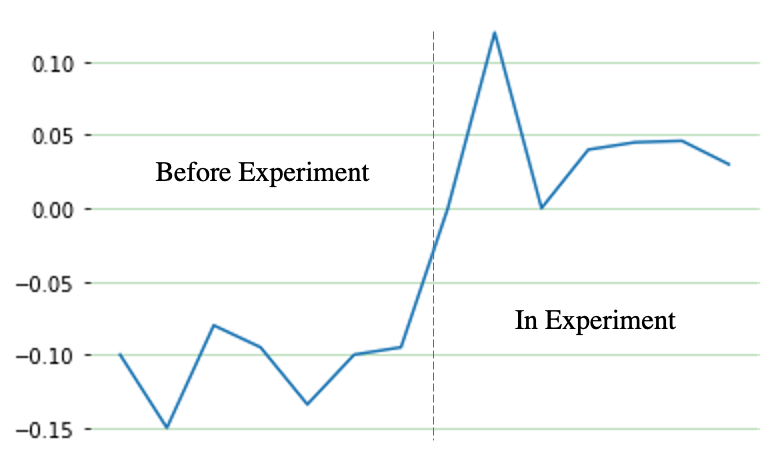}
  \caption{Live experiment results. On the x-axis is the date; on the y-axis is the relative difference in percentage between the experiment and control.}
  \label{fig:experiment_curve}
\end{figure}

To validate the effectiveness of the proposed matching framework named T2Diff, we conducted a week-long online A/B test (from March 27 to April 3, 2024) on a prominent large-scale short-video platform. The test involved over three million users, who were part of the experimental cohort. Within the experiment group, T2Diff is utilized as one of the potential candidate sources during the matching stage. We perform a comparison between the engagement rates of items selected from our source and those of other two-tower matching methods within the experiment. As evidenced in Table \ref{tab:online engagement}, the proposed approach exhibits superior performance over the other candidate sources. Furthermore, Figure \ref{fig:experiment_curve} demonstrates the live experiment results. On the x-axis is the date, and on the y-axis is the relative difference of a metric in percentage between the experiment and control. Relative to the control, the experiment group with T2Diff improves the average App usage duration by $\textbf{+0.143\%}$ with a $95\%$ confidence interval of $\textbf{(+0.02\%, +0.26\%)}$.

\begin{table}[h]
  \caption{A Comparison of Online Engagement Rates for Recommended Items between a regular two-tower matching method, DiffuRec and Our Approach: Analysis of Effective View Rate (EVR), Follow Rate (FTR), Average Played Duration (Play)}
  \label{tab_dataset}
  \begin{tabular}{ccccc}
    \toprule
    Metrics & EVR(\%) & FTR(\%) & Play(s) \\
    \midrule
    Regular Two-tower & 17.2\% & 0.45\% & 11.4s \\
    DiffuRec & \underline{21.9\%} & \underline{0.60\%} & \underline{15.26s} \\
    Ours & \textbf{24.6\%} & \textbf{0.67\%} & \textbf{20.96s} \\
    Improvement & +10.98\% & +11.67\% & +37.42\% \\
    \bottomrule
  \end{tabular}
  \label{tab:online engagement}
\end{table}

\section{Conclusion}
\label{sec:Conclusion}
In this paper, we propose a novel matching paradigm, T2Diff, which represents a significant advancement in generative cross-interaction decoupling architectures. This paradigm unleashes the potential of two-tower model by fetching the cross information between user and item representations which surmounts the challenge of "Late Interactions". T2Diff incorporates a generative module for precise reconstruction of the user's impending positive intention and introduces a mixed-attention mechanism to capture interactive signals based on the positive intention generated by the diffusion module. Moreover, the application of the diffusion model in matching stage to restore the target information offers expanded possibilities for generative retrieval methods. Extensive offline and online experiments demonstrate that T2Diff outperforms the SOTA two-tower retrieval models significantly, while numerous ablation studies validate the accuracy of our model design.

\clearpage
\bibliographystyle{ACM-Reference-Format}
\balance
\bibliography{sample-base}

\end{document}